\documentstyle[epsfig,12pt]{article}
\begin{document}

\newcommand \be  {\begin{equation}}
\newcommand \bea {\begin{eqnarray} \nonumber }
\newcommand \ee  {\end{equation}}
\newcommand \eea {\end{eqnarray}}

\title{{\bf ON A UNIVERSAL MECHANISM FOR LONG RANGED VOLATILITY CORRELATIONS}}

\author{Jean-Philippe Bouchaud$^{1,2}$, Irene Giardina$^3$ and Marc 
M\'ezard$^4$}

\date{\it
$^1$ Service de Physique de l'\'Etat Condens\'e,
 Centre d'\'etudes de Saclay, \\ Orme des Merisiers, 
91191 Gif-sur-Yvette Cedex, France \\ 
$^2$ Science \& Finance, 109-111 rue Victor-Hugo, 92532 France
\\
$^3$ Service de Physique Th\'eorique,
 Centre d'\'etudes de Saclay, \\ Orme des Merisiers, 
91191 Gif-sur-Yvette Cedex, France \\
$^4$ Laboratoire de Physique Th\'eorique et Mod\`eles Statistiques \\
Universit\'e Paris Sud, Bat. 100, 91 405 Orsay Cedex, France}
\maketitle

\begin{abstract}
We propose a general interpretation for long-range correlation 
effects in the activity and volatility of financial markets. This 
interpretation is
based on the fact that the choice between `active' and `inactive' strategies 
is subordinated to random-walk like processes. We numerically demonstrate our 
scenario 
in the framework of simplified market models, such as the Minority Game model 
with an 
inactive strategy. We show that real market data can be surprisingly
well accounted for by these simple models.
\end{abstract}

A well documented `stylized fact' of financial markets is volatility 
clustering 
\cite{volfluct1,volfluct2,MS,Book}. 
Figure 1 compares the 
time series of the daily returns of the Dow-Jones index since 
1900 and that of a Brownian random walk. Two features are immediately obvious 
to the eye: the volatility does indeed have rather strong 
intermittent fluctuations, and these fluctuations tend to persist in time. A 
more 
quantitative analysis shows that the
daily volatility $\sigma_t$ (defined, for example, as the average squared 
high 
frequency returns) has a log-normal
distribution \cite{stanley}, and that its temporal correlation function 
$\langle \sigma_t 
\sigma_{t+\tau} \rangle$ can be 
fitted by an inverse power of the lag $\tau$, with a rather small exponent in
the range $0.1 - 0.3$ \cite{volfluct2,PCB,stanley,muzy}. 
This suggests that there is no characteristic time scale
for volatility fluctuations: outbursts of market activity can persist for 
rather 
short times (say a
few days), but also for much longer times, months or even years. 
A very interesting observation is that these 
{\it long ranged} volatility correlations are observed on many different 
financial markets, with qualitatively similar
features: stocks, currencies, commodities or interest rates. This suggests 
that a common mechanism is at the
origin of this rather universal phenomenon.

\begin{figure}
\hspace*{+1cm}\epsfig{file=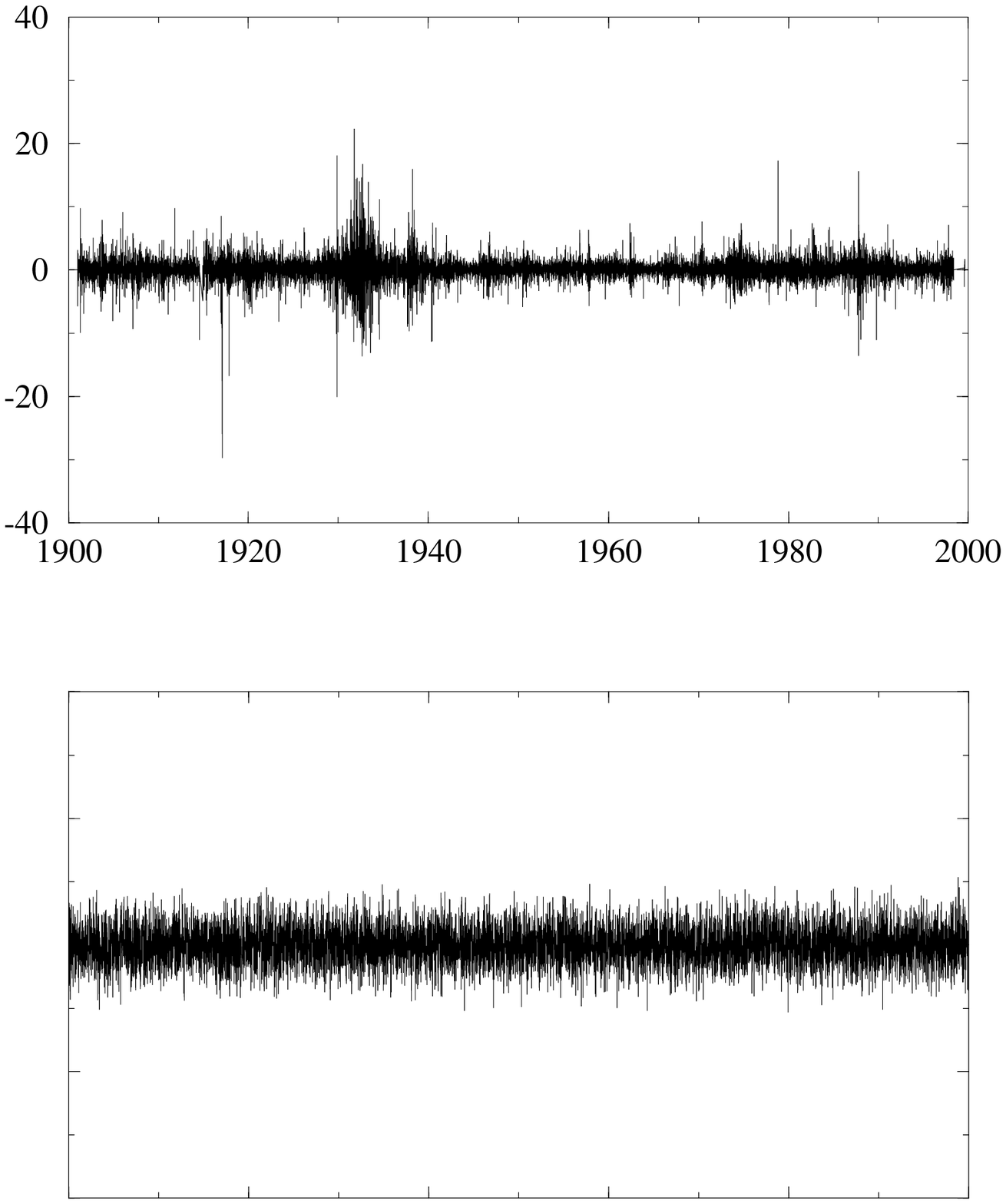,width=8cm}
\vskip 0.3cm \caption{\small Top panel: daily returns of the Dow-Jones index 
since 1900. 
One very 
clearly sees the intermittent nature of volatility fluctuations. Note in 
particular the 
sharp feature in the 30's. Lower panel: daily returns for a Brownian 
random walk,
showing a featureless pattern.
\label{fig1} }
\end{figure}

A first possibility is that the apparent lack of time scale associated 
with the power-law dependence of the
correlation function is a consequence of the fact that human activity is 
naturally rythmed by days, weeks,
months, quarters and years. Now, the ratio between these successive time 
scales is 
roughly constant. 
The superposition of correlation functions with time constants uniformly 
distributed on a log-scale may
easily be confused with a single power-law with a small exponent \cite{IOP}. 

However, very important insights into market dynamics have recently been 
gained 
by the study of several
agent based models \cite{Bak,MG1,CB,Farmer,Stauffer,Lux,Iori,farmerrev}. These models postulate some 
simple behaviour at the level of 
the agents and investigate 
the resulting price dynamics. Among others, the model by Lux and Marchesi 
\cite{Lux} assumes 
that each agent can 
behave, as a function of time, either as a fundamentalist (i.e. determining 
his 
action by comparing 
the current market price to some fundamental `true' price), or as a `trend 
follower', 
influenced by observed past trends on the price itself.
Agents switch between the two strategies as a function of their relative 
performance. 
Numerical 
simulations based on this model produce quite realistic price charts. 
In particular, long-ranged, power-law
type volatility correlations are reported. Another family of models, the 
`Minority Game' ({\sc mg}) and its variants \cite{MG1,MG2,Johnson,CGGS,Moro},
has recently become the focus of intense theoretical scrutiny. The Minority 
Game
describes the behaviour of competing agents that can choose between different 
individual strategies as
a function of their past performance. In its original version, this model is 
rather remote from financial
markets; in particular, there is no price dynamics. Several attempts have 
been 
made to generalize it and construct 
more realistic market models \cite{MG2}. As first noticed in \cite{Johnson}, 
if one allows 
the agents to be inactive,
intermittent volatility fluctuations can be generated. We have ourselves 
studied 
a market model that allows 
traders to switch between a bond market and a stock market, and accounts 
properly 
for their wealth 
balance and for market clearing. The phenomenology of this model is very rich 
and 
a detailed account of
our results will be published separately \cite{GBM}. One of our main result 
is the existence 
of a `turbulent' market
phase where volatility fluctuations are intermittent and show a power-law 
correlation 
(see Figure 3 below).
A very important point is that all these models are different in their 
details but 
all show qualitatively
similar behaviour, without the explicit introduction of any of the `human' 
time scales 
mentioned above. In other 
words, these agent based models assume a unique elementary time scale (say 
the `day') 
and the long-ranged 
volatility correlations spontaneously emerge from the dynamics.

We wish to propose a simple and robust mechanism to account for the 
appearance of 
these long-ranged
correlations in the above simplified models. We then argue that this 
mechanism 
also very naturally
operates in real financial markets, and accounts well for the empirical 
findings.
The following discussion is intentionally rather qualitative; more detailed 
and
technical results will be presented elsewhere \cite{GBM}.

The idea is the following: in the above models, scores are attributed by 
agents to their 
possible strategies,
as a function of their past performance. In a region of the parameter space 
where these  
models lead to an efficient market, the autocorrelation of the price 
increments is close 
to zero, which means that to a first 
approximation, no strategy can on average be profitable. This implies 
that for {\it any} reasonable definition of
the update of the scores, these scores will locally behave, as a function of 
time, 
as random walks. Furthermore, the scores 
associated to different strategies generically behave as {\it independent} 
random walks. Now, in all these 
models, the switch between two strategies occurs when their scores cross. 
Therefore,
in the case where each agent has two strategies, say one `active' (trading in 
the
market) and one `inactive' (holding bonds), the survival time of any one of 
these 
strategies will be given by the return time of a random walk (the difference 
between
the scores of the two strategies) to zero. The interesting point is that 
these return 
times are well known to be power-law distributed (see below):
this leads to the non trivial behaviour of the volume autocorrelation 
function. In other
words, the very fact that agents compare the performance of two strategies on 
a random signal 
leads to a multi-time scale situation.

More formally, let us define the quantity $\theta_i(t)$ that is equal to $1$ 
if agent $i$
is active at time $t$, and $0$ if inactive. The total activity is given by 
$a(t)=\sum_i 
\theta_i(t)$. The time autocorrelation of the activity is given by
\footnote{Up to an additive constant which disappears from the variogram.}:
\be
C_a(t,t')= \langle a(t) a(t') \rangle = \sum_{i,j} \langle 
\theta_i(t)\theta_j(t')
\rangle.
\ee
We will actually use in the following the so-called activity variogram, 
defined as:
\be
V_a(t,t')=\langle \left[a(t)-a(t')\right]^2 \rangle= 
C_a(t,t)+C_a(t',t')-2C_a(t,t').
\ee
One can consider two extreme cases which lead to the same result, up to a 
multiplicative constant: (a) agents follow 
completely different strategies and have independent activity patterns, i.e.
$\langle \theta_i\theta_j\rangle \propto \delta_{i,j}$ or (b) agents follow 
very 
similar strategies, for example by all comparing the perfomance of stocks to 
that of bonds,
in which case $\theta_i=\theta_j$. In both cases, one has $C_a(t,t')$ is 
proportional to $\langle \theta_i(t)\theta_i(t') \rangle$. This quantity can 
be 
computed in terms of the distribution $P(s)$ of the survival time $s$ of the 
strategies (in the following, we assume that both the inactive and active
strategies have the same survival time distribution).
Two cases must be distinguished: if $P(s)$ has a finite first moment $\langle 
s \rangle$ 
(finite average 
lifetime of the strategies), then $C_a(t,t')$ is stationary, i.e. it 
only depends on the difference $\tau=t'-t$. Introducing the Laplace 
transforms ${\cal L}C_a(E)$
and ${\cal L}P(E)$ of $C_a(\tau)$ and $P(s)$, the general 
relation between the two quantities reads \cite{GL}:
\be\label{result}
E{\cal L}C_a(E)=\left(1-\frac{2[1-{\cal L}P(E)]}{\langle s \rangle E
[1+{\cal L}P(E)]}
\right).
\ee
If one the other hand $P(s)$ has an infinite first moment, then $C_a(t,t')$ 
depends both 
on $t$ and 
$t'$: this is known as the {\it aging} phenomenon \cite{review,GL}. For an 
unconfined random
walk, the return time distribution decays as $s^{-3/2}$ for large $s$ 
and therefore its first moment is infinite. However, 
in all the models mentioned above, there exist `restoring' forces 
which effectively confine the scores to a finite interval \cite{GBM}. This 
can be attributed, 
both in 
the case of the {\sc mg} or of more realistic market models, to `market 
impact', which
means that good strategies tend to deteriorate because of their very use. 
There are many 
reasons to believe that such confining forces also operate in financial 
markets. 
The consequence of
these effects is to truncate the $s^{-3/2}$ tail for values of $s$ larger 
than a certain 
equilibrium time $s_0$. Therefore, the first moment of $P(s)$ actually 
exists, such that 
Eq. (\ref{result}) is valid. Nevertheless, one can see
from Eq. (\ref{result}) that the characteristic 
$s^{-3/2}$ behaviour of $P(s)$ for short time scales leads to $E{\cal 
L}C_a(E) \sim
1 - B/\sqrt{E} + ...$ for $s_0^{-1} \ll E \ll 1$. This in turn leads to a 
singular behaviour
for the variogram $V_a(\tau)$ at small $\tau$'s, as $V_a(\tau) \propto 
\sqrt{\tau}$, 
before saturating to a finite value for $\tau \sim s_0$. Intuitively, this 
means that 
the probability for the activity to have changed significantly between $t$ 
and $t+\tau$ is
proportional to $\int_0^\tau ds \ s P(s) \propto \sqrt{\tau}$ (for $\tau \ll 
s_0$), where 
$s P(s)$ is the probability to be at time $t$ playing a strategy with 
lifetime $s$.

\begin{figure}
\hspace*{+1cm}\epsfig{file=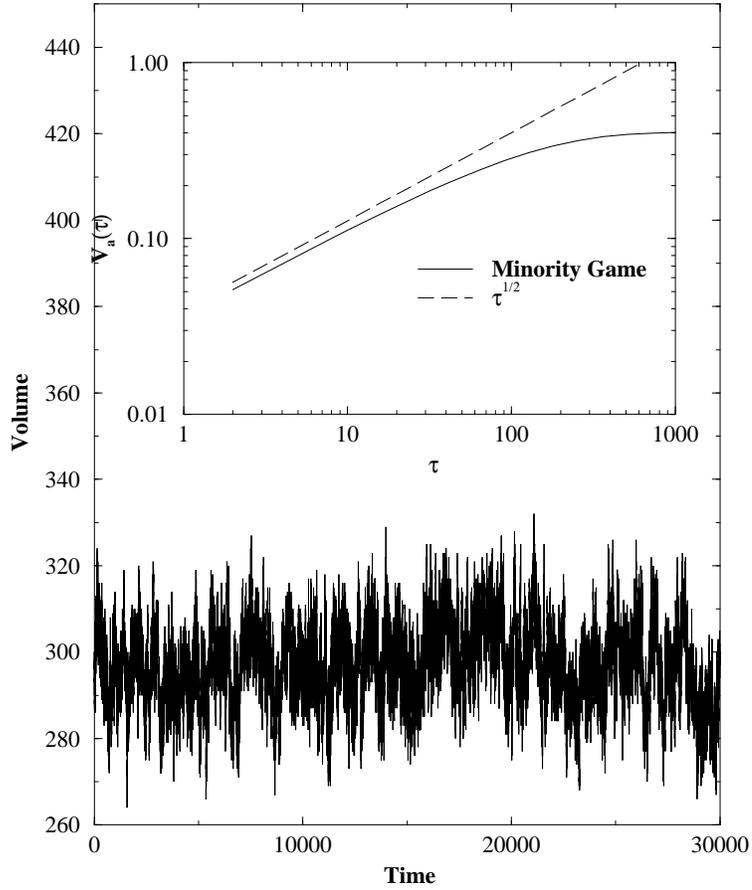,width=8cm}
\vskip 0.3cm \caption{\small Volume of activity (number of active agents) as 
a function of time 
in the
{\sc mg} with two active strategies and one inactive strategy per agent, 
and 
for $\alpha=0.51$ ($\alpha_c \simeq 1.$). The number of agents is $501$. 
Inset: The corresponding 
activity variogram 
as a function of the lag $\tau$, in a log-log plot to emphasize the 
$\protect\sqrt{\tau}$
singularity at small $\tau$'s.
\label{fig3} }
\end{figure}

Let us illustrate this general scenario with the example of the {\sc mg} 
with an inactive
strategy, first introduced in \cite{Johnson}. Each agent has a certain number 
of fixed strategies
to choose from. A strategy is a mapping from a signal (for example the past 
history) into a 
decision, say $+1$ or $-1$. The aim of the game at each time step is to 
make the decision 
that is chosen by the minority of the agents at that time \cite{MG1}. 
If a strategy is successful (or
would have been if it had been played), its score increases, conversely, 
if the wrong decision is made (or again,
would have been if it had been played), the score decreases. The
chosen strategy is the one that has the highest score. If all the strategies 
of an agent 
have negative scores, then the agent does not play. The relevant parameter 
$\alpha$ of this 
model
is the ratio of the number of possible histories to the number $N$ of agents.
The history is given by the $M$ past steps of the game, therefore 
$\alpha=2^M/N$.
One finds 
\cite{Canat,GBM} that
there is a critical value $\alpha_c$ above which all agents finally become 
inactive. 
Below this value, the activity is non zero. A plot of the activity as a 
function of time in
this model is given in Figure 2, for a value of $\alpha$ smaller 
than $\alpha_c$.
In the inset, we have plotted the activity variogram $V_a(\tau)$, which reveal
the characteristic $\sqrt{\tau}$ singularity discussed above, before 
saturating for
large $\tau$ ($\sim s_0$). 
This $\sqrt{\tau}$ singularity is present in the whole active phase $\alpha < 
\alpha_c$,
although $s_0$ is large compared to $1$ only if $\alpha$ is not too 
small. Very similar 
variograms 
have also been found in the more realistic market
model that we have investigated, showing the universality of this result
(see Figure 3) \cite{GBM}. Note that a 
similar mechanism might also be present in
the Lux-Marchesi model, where it has been observed that the activity bursts 
are 
associated to a large number of agents switching from being `fundamentalists' 
to being 
`trend followers' \cite{Lux}.

\begin{figure}
\hspace*{+1cm}\epsfig{file=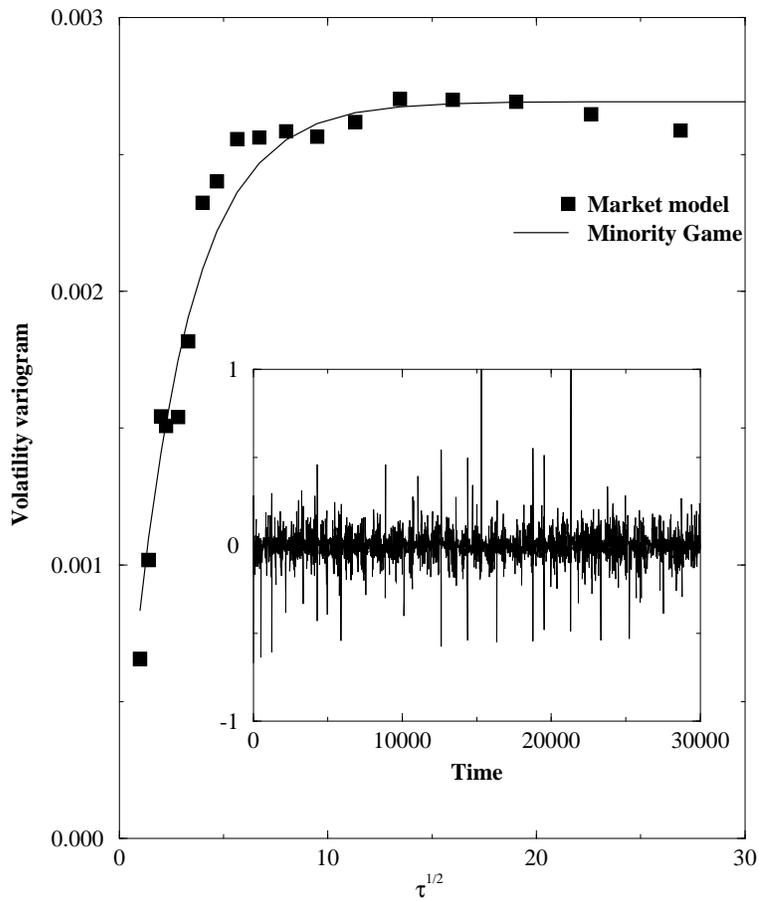,width=8cm}
\vskip 0.3cm \caption{\small Volatility variogram (squares) for a market 
model, 
inspired from the {\sc mg}, where agents buy and sell a stock, and
can switch from the stock market to bonds. The full line corresponds to the 
prediction of the
simple {\sc mg} with an inactive strategy. 
Inset: Price changes as a function of time in the same model,
showing volatility clustering.
\label{fig2} }
\end{figure}

It is interesting to compare the above results with real market data. Figure 
4 shows the 
volume of activity on the S\&P 500 futures contract in the years 1985-1998. 
This plot 
is to the eye very similar to the one of Figure 3, obtained with the {\sc mg}.  
This is quantitatively confirmed by the activity variogram, shown in the 
inset. 
On the same graph, we have
reproduced the {\sc mg} result. Both the time scale and the volume scale
(arbitrary in the {\sc mg} model) have been adjusted to get the best 
agreement.
Furthermore, a constant has been added to $V_a$ (corresponding to a 
$\delta_{\tau,0}$ contribution to $C_a$), to account for the fact that part 
of 
the trading activity is
certainly white noise (e.g. motivated by news, or by other non strategic 
causes). 
As can be seen, the agreement is rather good. Most significant is the 
clear $\sqrt{\tau}$ behaviour at small $\tau$ ($\tau < 50$ days). We therefore
suggest that the effect captured by the {\sc mg} (Figure 2) or more 
sophisticated 
variants (Figure 3), 
namely the subordination of the activity on random walk like signals, is also 
present in 
real markets. It seems to us that this makes perfect sense since market 
participants
indeed compare the results of different strategies to decide 
whether they should
remain active in a market or leave it. 

\begin{figure}
\hspace*{+1cm}\epsfig{file=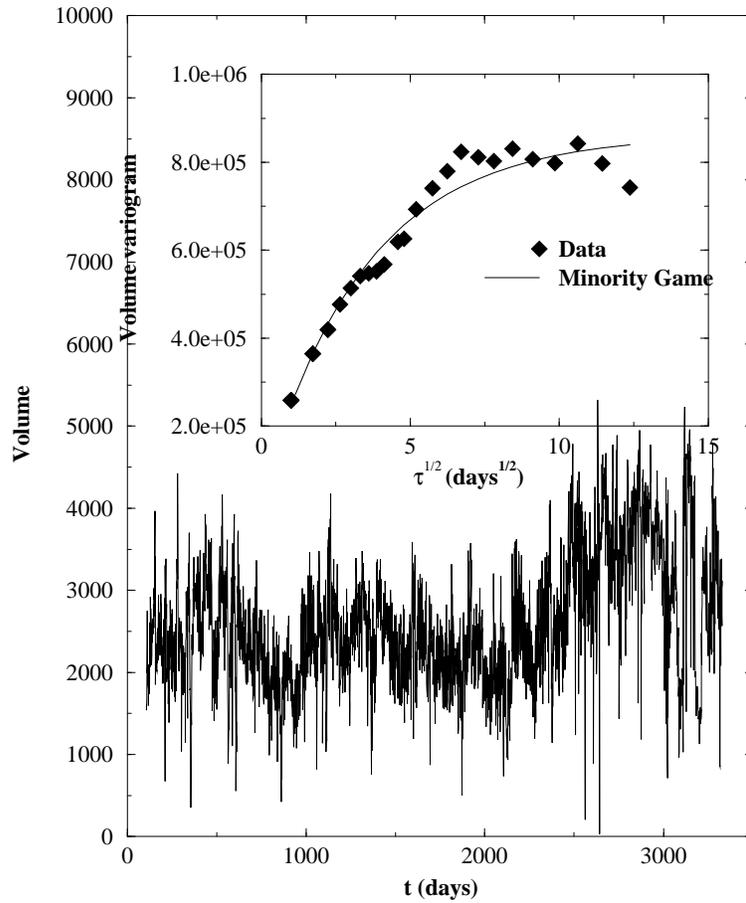,width=8cm}
\vskip 0.3cm \caption{\small Total daily volume of activity (number of trades) on 
the S\&P 500 futures contracts in the years 1985-1998: compare to Figure 2. Inset: 
Corresponding variogram 
(diamonds) 
as a function
of the square-root of the lag. Note the clear linear behaviour for small $
\protect\sqrt{\tau}$. The full line is the {\sc mg} result, with both 
axis
rescaled and a constant added to account for the presence of `white noise' 
trading.
\label{fig4} }
\end{figure}

Since the volatility and the volume of 
activity are
strongly correlated in financial markets \cite{Bonnano,Gopi}, our 
interpretation should naturally carry over to volatility fluctuations 
as well. This is illustrated in Figure 5, where the variogram of the log-volatility for major stock indices is shown, together with the very same
{\sc mg} result. Again, the agreement is very good. We have also shown
for comparison the prediction of the multifractal model of ref. \cite{muzy},
$V_a(\tau)=2\lambda^2 \log(\tau/\tau_0)$. It is interesting to note that 
the two model, although very different, lead to nearly indistinguishable 
numerical fits.

\begin{figure}
\hspace*{+1cm}\epsfig{file=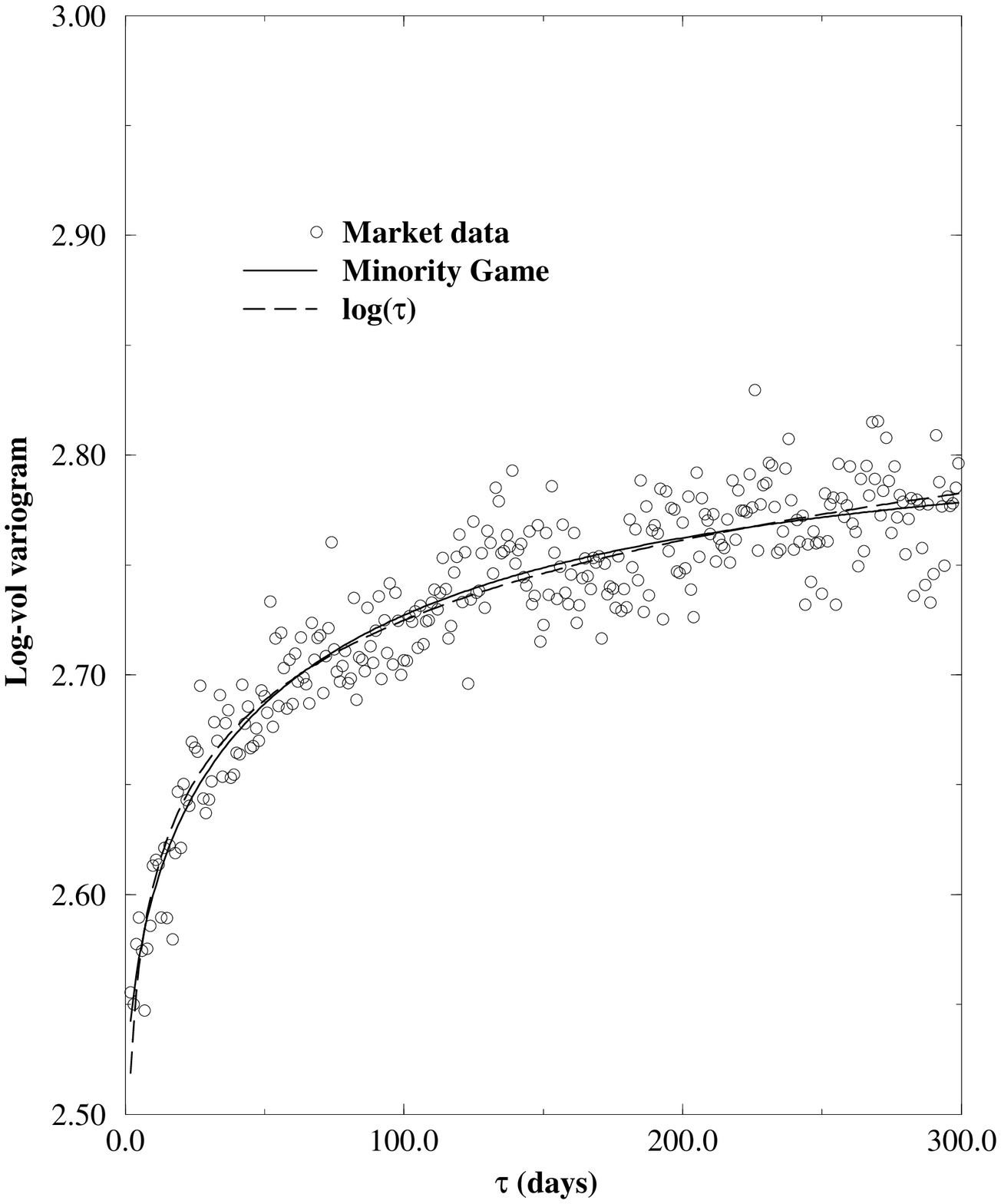,width=8cm}
\vskip 0.3cm \caption{\small Variogram of the log-volatility, $\langle
\log^2(\sigma_t/\sigma_{t+\tau})\rangle$ as a function of $\tau$, averaged over 17 different 
stock indices (American, European, Asian). The full line is the {\sc mg} result, with again both 
axis
rescaled and a constant added to account for the presence of `white noise' 
trading. The dashed line is the prediction of the multifractal model of
\protect\cite{muzy}, and is nearly indistinguishable from the {\sc mg} result.
\label{fig5} }
\end{figure}

The analogy between volatility clustering in financial markets and 
intermittency effects
in turbulent flows has recently been emphasized \cite{Ghas,muzy,BPM}. It is 
tempting to speculate
that the mechanism discussed here might also be at work in turbulent flows, 
where outburts
of activity are due to localized structures \cite{Frisch}. If the motion of 
these localized 
structures locally resembles that of a random walk, similar conclusions can 
be expected.

In summary, we have proposed a very general interpretation for long-range 
correlation 
effects in the activity and volatility of financial markets. This 
interpretation is
based on the fact that the choice between different strategies is 
subordinated to random-walk 
like processes. We have numerically demonstrated our scenario in the 
framework of simplified 
market models, and showed that, somewhat surprisingly, real market data can 
actually be quite accurately accounted for by these simple models (see Figs 4 
and 5). 
\vskip 1cm

{\it Acknowledgements} We wish to thank M. Potters for several important discussion on this topic,
and A. Matacz and Ph. Seager for providing the data of Figs 4 and 5, and for
interesting remarks.
Useful interactions with G. Canat, A. Cavagna, D. Farmer, E. Moro, J.P. Garrahan, N. 
Johnson, M.
Marsili,  D. Sherrington and H. Zytnicki are acknowledged.

\end{document}